\documentclass[lettersize,journal]{IEEEtran}
\usepackage{amsmath,amsfonts}
\usepackage{algorithmic}
\usepackage{algorithm}
\usepackage{array}
\usepackage[caption=false,font=normalsize,labelfont=sf,textfont=sf]{subfig}
\usepackage{textcomp}
\usepackage{stfloats}
\usepackage{url}
\usepackage{verbatim}
\usepackage{graphicx}
\usepackage{cite}
\usepackage{hyperref}
\begin{document}

\title{Low-cost Microfluidic Testbed for Molecular Communications with Integrated Hydrodynamic Gating and Screen-printed Sensors}

\author{Maide Miray Albay$^{\ast}$, Eren Akyol$^{\ast}$, Fariborz Mirlou$^{\ast}$, Levent Beker, and Murat Kuscu
\thanks{*These authors contributed equally.}
        \thanks{M. M. Albay, F. Mirlou, and L. Beker are with the Bio-integrated Microdevices Lab (BMDL), Koç University, Istanbul, Turkey (e-mail: \{malbay23, fmirlou20, lbeker\}@ku.edu.tr).}
        \thanks{E. Akyol and M. Kuscu are with the Nano/Bio/Physical Information and Communications Laboratory (CALICO Lab) and Center for neXt-generation Communications (CXC), Koç University, Istanbul, Turkey (e-mail: \{eakyol22, mkuscu\}@ku.edu.tr).}
	   \thanks{This work was supported in part by The Scientific and Technological Research Council of Turkey (TUBITAK) under Grants \#120E301, \#123E516, \#118C295.}
}

\maketitle
\begin{abstract}

Molecular Communications (MC), transferring information via chemical signals, holds promise for transformative healthcare applications within the Internet of Bio-Nano Things (IoBNT) framework. Despite promising advances toward practical MC systems, progress has been constrained by experimental testbeds that are costly, difficult to customize, and require labor-intensive fabrication. Here, we address these challenges by introducing a low-cost ($\sim$\$1 per unit), rapidly fabricated ($<$1 hour), and highly customizable microfluidic testbed that integrates hydrodynamic gating and screen-printed potentiometric sensors. This platform enables precise spatiotemporal control over chemical signals and supports reconfigurable channel architectures along with on-demand sensor functionalization. As a proof of concept, we demonstrate a pH-based MC system combining a polyaniline (PANI)-functionalized sensor for real-time signal detection with a programmable hydrodynamic gating architecture, patterned in a double-sided adhesive tape, as the transmitter. By dynamically mixing phosphate-buffered saline (PBS) with an acidic solution (pH 3), the testbed reliably generates pH-encoded pulses. Experimental results confirm robust control over pulse amplitude and pulse width, enabling the simulation of end-to-end MC scenarios with 4-ary concentration shift keying (CSK) modulation. By combining affordability and rapid prototyping without compromising customizability, this platform is poised to accelerate the translation of MC concepts into practical IoBNT applications.

\end{abstract}


\section{Introduction}
\IEEEPARstart{M}{olecular} Communications (MC) is a bio-inspired communication paradigm  where information is encoded, transmitted, and received via the release and sensing of molecules, promising particularly for disruptive intra-body healthcare applications within IoBNT \cite{akyildiz2015internet}. Building on initial theoretical foundations exploring channel capacity, modulation, detection, and coding \cite{kuscu2019transmitter}, MC has recently advanced toward practical applications. A key driver of this shift has been experimental MC testbeds \cite{lotter2301experimental,lotter2023experimental}. 

Several MC testbeds have been developed across various physical scales, from macro-scale benchtop setups to microfluidic platforms, using diverse channel media, molecular messengers, and transmitter-receiver architectures \cite{angerbauer2023salinity, bhattacharjee2020testbed, walter2023real, kuscu2019transmitter}. Early prototypes often used off-the-shelf components, such as programmable sprayers and syringes as transmitters, polymer tubings and glass vessels as channels, and alcohol sensors and pH probes as receivers \cite{lotter2301experimental}. While enabling rapid prototyping, these setups limited scalability, customization, and communication schemes that could be studied. Recent work, focusing on micro/nanoscale testbeds, has utilized micro/nanofabrication methods and advanced materials (e.g., graphene, PDMS, nanoparticles) to construct custom-designed transmitters, channels, and receivers \cite{kuscu2021fabrication, angerbauer2023salinity, naqvi2023testbed, civas2023graphene}. These testbeds often incorporate microfluidics for precise control over channel geometry and flow, replicating realistic application environments \cite{hamidovic2024microfluidic}. However, specialized equipment (e.g., cleanroom facilities) and costly fabrication protocols limit widespread adoption.

\begin{figure*}[t]
    \centering
    \includegraphics[width=1\textwidth]{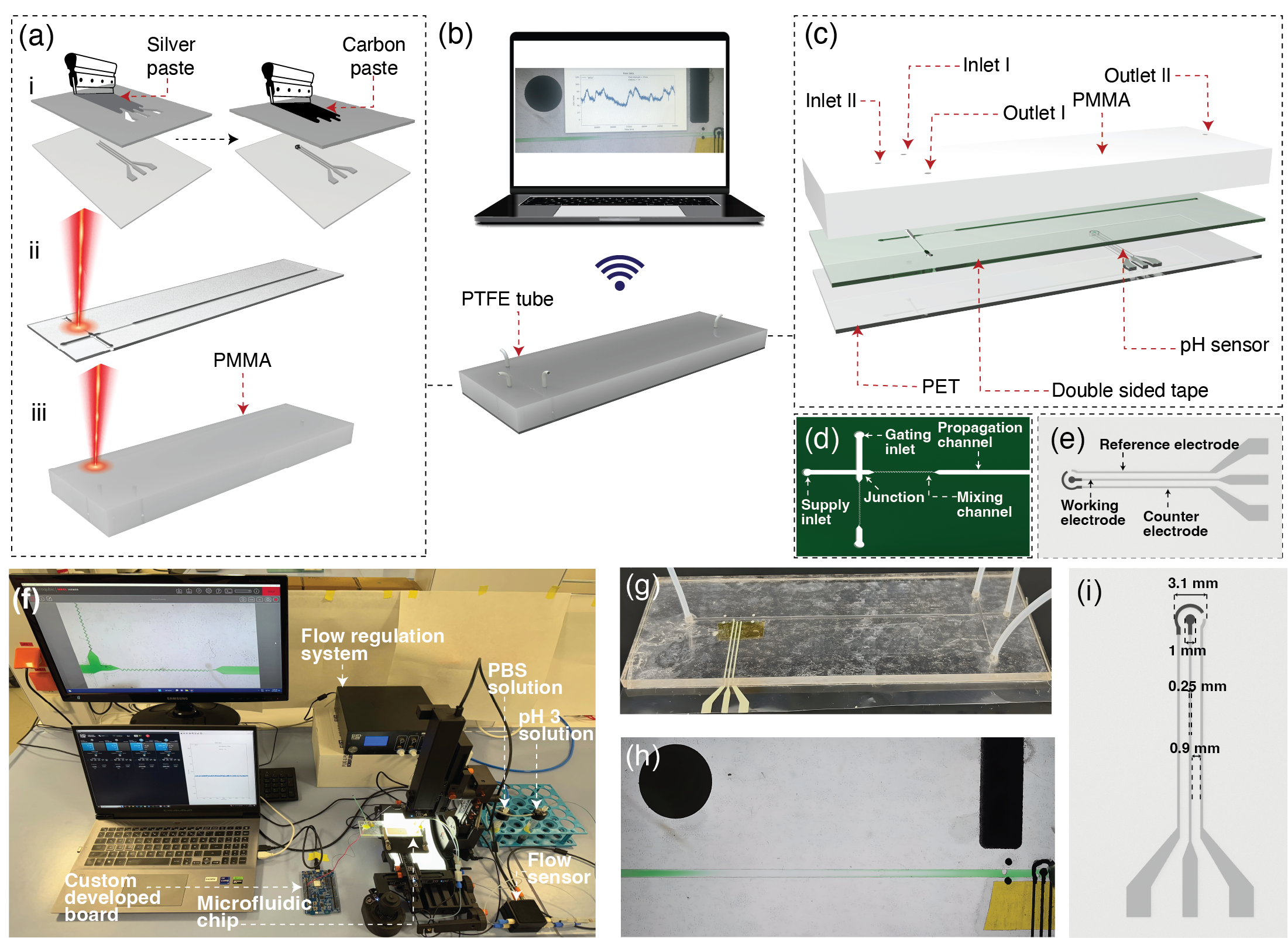} 
    \caption{(a) Fabrication flow of the microfluidic chip: i) Fabrication of screen-printed electrodes. ii) Fabrication of hydrodynamic gating pattern on double sided tape. iii) Fabrication of inlet and outlet on the PMMA. (b) Continuous monitoring of pH changes and data acquisition using a BLE-integrated custom-developed circuit board. (c) Exploded view of microfluidic chip. (d) Close-up schematic view of the hydrodynamic gating-based transmitter architecture. (e) Schematic view of screen-printed pH sensor. (f) Experimental setup. (g) Real image of microfluidic channel. (h) Pulse generation with hydrodynamic gating transmitter. (i) Dimensions of the screen-printed sensor.}
    \label{fig:setup}
\end{figure*}

In this paper, we introduce a cost-effective ($\sim$\$1 per unit), microfluidic MC testbed that requires only widely available materials and simple fabrication methods (Fig. \ref{fig:setup}(a-c)), yet offers extensive customization to study a broad range of MC scenarios at the micro-scale. Central to the testbed is hydrodynamic gating-based pulse-shaping functionality (Fig. \ref{fig:setup}(d)), introduced in our recent work \cite{bolhassan2024microfluidic, akyol2024experimental}, enabling precise MC signal generation in microfluidic channels. The integrated receiver architecture incorporates screen-printed electrodes (Fig. \ref{fig:setup}(e)) that can be rapidly functionalized for electrochemical sensing of various types of molecular messengers. The end-to-end MC testbed is readily compatible with commonly available laboratory equipment, such as pressure-based flow regulators, syringe pumps, and microcontrollers (Fig. \ref{fig:setup}(f)).

To demonstrate the testbed’s capabilities, we implemented a pH-based MC system. pH modulation is biologically relevant in processes like wound healing and cellular metabolism, driving interest in pH as an information transfer modality for IoBNT \cite{walter2023real,farsad2016molecular}. We fabricated a custom-designed, screen-printed, polyaniline (PANI)-coated potentiometric pH sensor on a polyethylene terephthalate (PET) substrate as the receiver. Meanwhile, a hydrodynamic gating architecture, patterned on double-sided tape with a poly(methyl methacrylate) (PMMA) cover, constituted the transmitter, enabling precise pH modulation (Fig.\ref{fig:setup}(g)). 
After confirming the receiver’s performance, we conducted microfluidic experiments to evaluate the transmitter’s capacity to generate precise pH-based MC signals (Fig. \ref{fig:setup}(h)). PBS (pH 7.4) served as a baseline, and an acidic solution (pH 3) was introduced at controlled flow rates to assess amplitude modulation and verify the transmitter’s reliability in maintaining or shifting pH levels. We also examined pulse width modulation, confirming consistent pulse profiles with programmable amplitudes and widths. Finally, end-to-end MC was demonstrated using a 4-ary concentration shift keying (CSK) modulation scheme with varying pulse widths and inter-pulse durations, including cases with and without inter-symbol interference (ISI). These experiments show the testbed's versatility and capacity to replicate diverse MC scenarios. Although focused on pH-based MC, this methodology can extend to ion-based or other molecular signaling modalities if chemical agents are available to functionalize the receiver's screen-printed electrodes.

\section{Materials and Methods}
\subsection{Sensor Preparation}
\begin{figure}[h]
    \centering
    \includegraphics[width=\columnwidth]{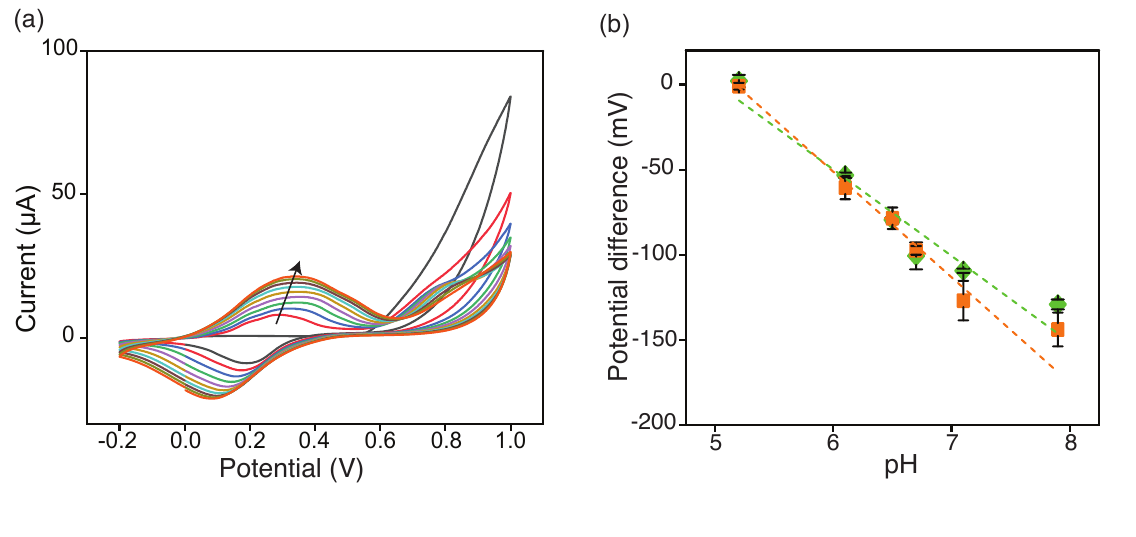} 
    \caption{(a) Cylic voltammetry of PANI deposition on fabricated pH sensor. (b) Calibration curves of pH sensors.}
    \label{fig:PANI} 
\end{figure}

Potentiometric pH sensors, a 3-electrode system with carbon working and counter electrodes and a silver reference electrode, were fabricated via screen-printing on a commercial PET substrate. Silver and carbon are common materials in screen-printed electrodes, with silver valued for its conductivity and carbon for cost-effectiveness and chemical stability \cite{franco2020screen}. Stencil masks were designed and patterned on 40 $\mu$m PET films using AutoCAD and an IR laser machine. Separate masks were developed for each screen-printing step. First, silver paste applied to pattern the reference electrode and contact pad. Then, the working and counter electrodes were patterned with carbon paste. After each application, the paste was cured at $70$ $^\circ$C for 30 minutes. Once fabricated to the dimensions in Fig. \ref{fig:setup}(i), the sensor was functionalized with PANI in a solution of $0.1$ M aniline in $1$ M HCl via cyclic voltammetry from $-0.2$ to $1$ V for $10$ cycles (Fig. \ref{fig:PANI}(a)) at a scan rate of $50$ mV/s. The PANI-coated sensors were rinsed with deionized water and dried with $N_2$ gas before use.

The electropolymerization of aniline on the pH sensor's working electrode shows redox peaks, confirming PANI film formation. The initial oxidation peak (between +0.8 and +0.9 V vs. reference in acidic environments) indicates aniline radical cation generation, initiating polymer growth. As the film develops, additional peaks reflect PANI oxidation state transitions. These peaks become more distinct and shift slightly in potential as the film thickness and conductivity increases. During polymerization, the yellowish solution darkens to green, confirming successful electropolymerization \cite{gvozdenovic2011electrochemical}.

Sensor sensitivity was tested by measuring potential changes across different pH levels. Potentiometric measurements between the PANI-coated working electrode and the reference electrode were conducted using the standard open circuit potential (OCP) technique via the electronic circuit board. Sensor characterization involved adjusting pH with $0.1$ M NaOH (basic) and $0.1$ M HCl (acidic) solutions. For calibration, $100$ $\mu$L of solutions with varying pH (5.2 to 7.9) were applied to the sensor, and potentiometric measurements were performed. As illustrated in \ref{fig:PANI}(b) the calibration curves exhibit a near-linear voltage response of 65 mV per pH unit.

\begin{figure}[t]
    \centering
    \includegraphics[width=\columnwidth]{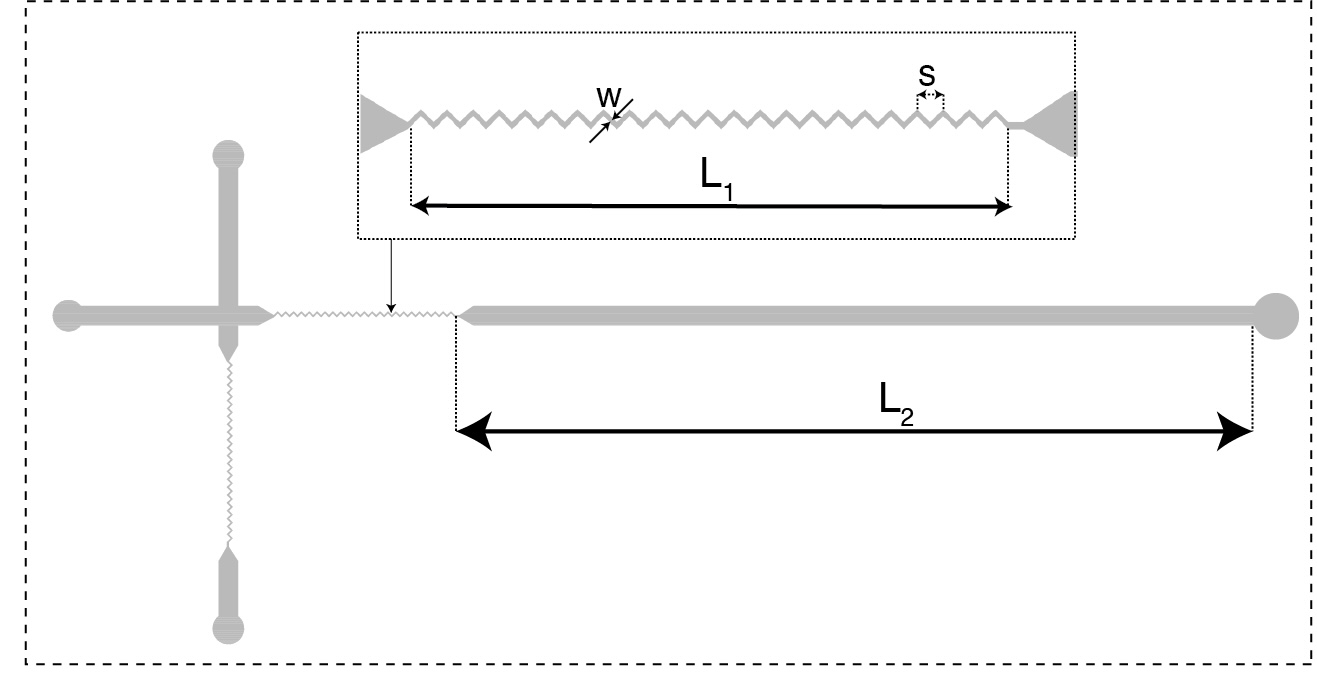} 
\caption{Hydrodynamic gating-based transmitter design with zigzag mixing structure. L\textsubscript{1}= 9.2 mm, L\textsubscript{2}= 80.5 mm, w= 100 $\mu$m, s= 400 $\mu$m.}
    \label{fig:design} 
\end{figure}

\subsection{Potentiometric Readout Circuit}

A custom-designed circuit board with a voltage divider was developed for potentiometric pH sensor measurements. The circuit captures analog voltage signals from the sensor and converts them to digital using the microcontroller's 12-bit ADC. Processed data are then transmitted to personal smart devices via Bluetooth Low Energy (BLE) for further analysis. The analog signal conditioning path includes a programmable gain amplifier (PGA) set at unity gain ($\times1$) and an infinite impulse response (IIR) filter with a $1$ Hz cutoff frequency to reduce noise and interference. Toothless alligator clips were used to securely connect the flexible sensor to the circuit board.

\subsection{Preparation of Microfluidic Chip with Hydrodynamic Gating Architecture and Device Integration}
\label{gating}

The microfluidic transmitter uses hydrodynamic gating to control fluid flow in a microchannel by dynamically adjusting the flow rate of intersecting streams, enabling precise pulse generation \cite{akyol2024experimental,bolhassan2024microfluidic}. Having a cross-shaped design, it includes a mixing structure between the junction and propagation channel, as shown in Fig. \ref{fig:setup}(d) and Fig. \ref{fig:design}. It operates by switching between ON and OFF modes, adjusting the relative flow rates of supply (pH 3) and gating (PBS, pH 7.4) solutions to regulate supply solution entry into the mixing region. During gating ON, the gating inlet flow dominates, preventing supply solution entry into the propagation channel and defining the inter-pulse duration, $T\textsubscript{pi}$. During gating OFF, the supply solution enters the mixing structure as the gating inlet flow rate is reduced, determining the gating duration, $T_\mathrm{g}$. After generating the desired pulse profile, the mixing region ensures homogeneous cross-sectional mixing of solutions in the propagation channel.

\begin{figure}[b]
    \centering
    \includegraphics[width=0.8\columnwidth]{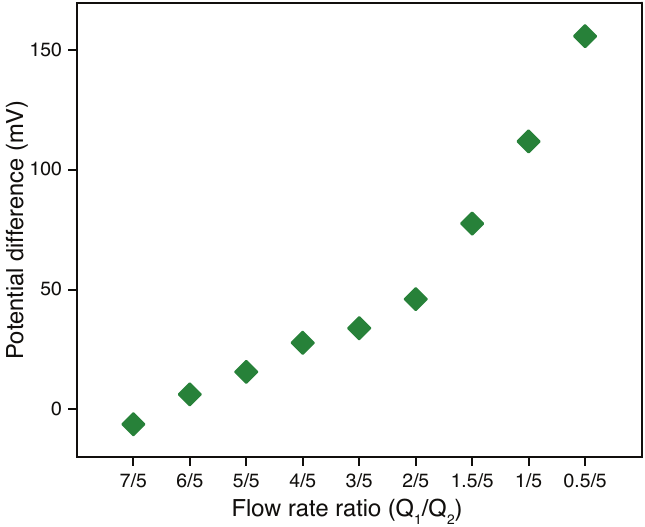} 
    \caption{Receiver response as a function of the flow rate ratio between pH 3 and PBS solutions corresponding to different pH levels for amplitude modulation.}
    \label{fig:amplitude} 
\end{figure}

A microfluidic chip with hydrodynamic gating/passive mixing structure and the propagation channel was designed in AutoCAD. An IR laser patterned double-sided tape (TESA SE, Germany), while a CO$_2$ laser (D960, Almira Laser, Türkiye) structured PMMA for inlets and outlets. The patterned layers were precisely aligned, and PTFE tubes were inserted and sealed with Loctite (Henkel, Germany) epoxy to prevent leakage, completing device integration (Fig. \ref{fig:setup}(g)).

\section{Results}

\subsection{Amplitude Modulation}


To determine the range and steps of the amplitude modulation, we analyzed the stationary sensor response by filling the propagation channel with solutions of varying pH levels. For this analysis, we adjusted the flow rate of the gating solution between $0.5$ and $7 \mu\text{L/min}$ while the flow rate of the supply solution was fixed at $5 \mu\text{L/min}$, thereby controlling the proportion of supply and gating solutions in the propagating mixture. Each pair of flow rates was held constant for two minutes to obtain a uniform concentration in the propagation channel and stabilize the sensor response. As a result, even small adjustments in the flow rate ratio, lead to clearly distinguishable levels in the potential difference, as depicted in Fig. \ref{fig:amplitude}. Moreover decreasing the flow rate ratio results in increased proportion of the supply solution in propagating mixture, which in turn lowers the pH level. The observed monotonic increase in the potential difference as a function of flow rate ratio, validates the high precision and programmability of the hydrodynamic gating-based transmitter as well as the sensor's high sensitivity.

\subsection{Pulse Shaping} 

The microfluidic transmitter generates programmable pulse profiles by employing the hydrodynamic gating technique in which a pulse is generated through a cycle of gating OFF and ON modes. As discussed in \ref{gating}, $T_\mathrm{g}$ and $T\textsubscript{pi}$ are determined by the duration of the gating modes and the pulse amplitude can be adjusted by varying the flow rate of the gating solution during the OFF mode, which controls the proportion of the supply and gating solution entering into the propagation channel. This precise pulse shaping technique enables programmable pulses with controlled spacing, width, and amplitude, while the integrated mixing structure ensures uniform cross-sectional distribution \cite{akyol2024experimental}.

\begin{figure*}[t]
    \centering
    \includegraphics[width=\textwidth]{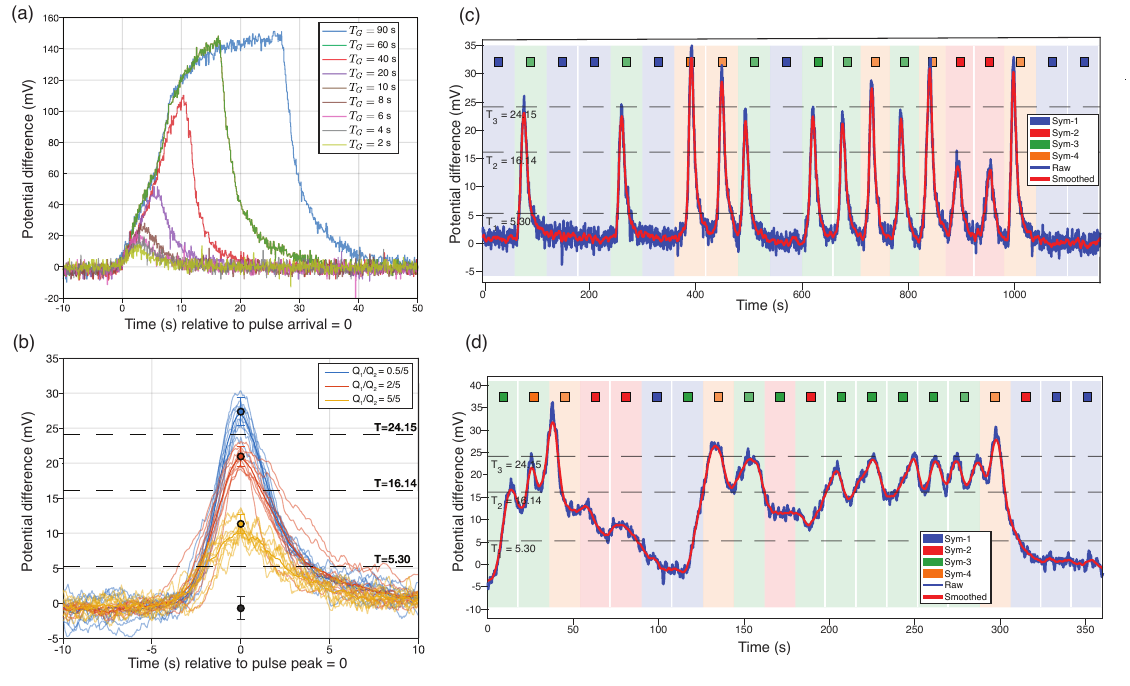} 
    \caption{(a) Receiver responses to four modulation levels (\(Q_1/Q_2 = 0.5/5\), \(2/5\), \(5/5\), and \(7/5\)) and thresholds (\(T_1 = 6.30\,\mathrm{mV}\), \(T_2 = 16.14\,\mathrm{mV}\), and \(T_3 = 24.15\,\mathrm{mV}\)), with $T_\mathrm{g}$ held constant at 20\,s. (b) Pulse width characterization for different \(T_\mathrm{g}\) ranging between 90\,s and 2\,s, with \(Q_1/Q_2 = 0.5/5\). (c) Receiver response to a randomly generated sequence of 20 symbols without ISI, \(T_\mathrm{g} = 10\,\mathrm{s}, T_\text{pi} = 20\,\mathrm{s}\). (d) Receiver response to a randomly generated sequence of 20 symbols under ISI effects, \(T_\mathrm{g} = 3\,\mathrm{s}, T_\text{pi} = 10\,\mathrm{s}\).}
    \label{fig:pwandcomm} 
\end{figure*}
\subsubsection{Pulse Width Characterization}

To evaluate the pulse shaping performance and characterize the sensor response to different pulse profiles, we conducted experiments by varying $T_\mathrm{g}$ to modulate pulse width and adjusting the flow-rate ratio to tune pulse amplitude. To characterize the pulse width, we evaluated varying $T_\mathrm{g}$ values ranging from $90$ s to $2$ s, maintaining a constant flow rate ratio of $0.5/5$. This particular ratio was chosen as it generates the highest potential difference within the modulation range, as demonstrated in Fig. \ref{fig:amplitude}, ensuring detectable peaks even during the transmission of pulses with smaller widths. As shown in Fig. \ref{fig:pwandcomm}(a), decreasing $T_\mathrm{g}$ leads to a reduction in pulse width, since a shorter $T_\mathrm{g}$ limits the duration of the OFF mode, allowing less time for the supply solution to enter the propagation channel. Moreover, longer pulses (e.g. $90$ s, $60$ s) produce responses with higher peaks but introduced significant time overhead, while shorter pulses (e.g. $4$ s and $2$ s) yield closer peaks to the baseline, making them potentially indistinguishable from the sensor response to the gating solution.

\subsubsection{Pulse Amplitude Characterization}

To characterize the pulse amplitude, we evaluated four different flow rate ratios—$Q_1/Q_2 = 0.5/5$, $2/5$, $5/5$, and $7/5$—while holding $T_\mathrm{g}$ constant at $10$ s, and for each ratio, nine transmitted pulses were analyzed. Particularly, the $7/5$ ratio corresponds to the transmission of the gating solution. Fig. \ref{fig:pwandcomm}(b) illustrates the receiver responses as potential differences across the four modulation levels, where thin lines represent individual pulse responses and bold lines indicate the average response for each level. Notably, increasing the flow rate ratio reduces the proportion of the supply solution in the generated pulse, leading to smaller peaks. Moreover, the minimal amplitude variation within each modulation level indicates the precise and programmable amplitude modulation of the pulses, as well as the receiver's sensitivity in detection.

\subsection{Communication Experiments}

After verifying the testbed's capacity to generate and detect programmable MC signals, we conducted pH-based MC experiments to illustrate its applicability in multi-level symbol transmission.  
Specifically, we selected four distinct pH levels, corresponding to flow ratios of \(Q_1/Q_2 = 0.5/5\), \(2/5\), \(5/5\), and \(7/5\), Among these, the ratio \(7/5\) served as the baseline level (PBS), effectively functioning as the gating ON condition. The chosen levels provide clearly separated amplitude responses, making them suitable for symbol encoding. 

During the experiments, pseudo-random sequences of four symbols were generated with the transmitter introducing PBS for a duration of \( T_\mathrm{pi} \) between each symbol. Receiver responses, thresholds, and classifications were then assessed based on the maximum potential difference of the smoothed data. The results for the initial $20$ transmissions under two different ISI conditions are presented in Fig. \ref{fig:pwandcomm}(c-d).

In Fig. \ref{fig:pwandcomm}(c), we observe the receiver response with minimal ISI, achieved by setting $T_\mathrm{g}$ = 10\,s and $T\textsubscript{pi}$ = 20\,s. This ensures sufficient spacing between pulses, allowing them to preserve their original amplitude and width. The results revealed clear threshold crossings at \( T_1 = 6.30\,\mathrm{mV}\text{,  } T_2 = 16.14\,\mathrm{mV}, \text{ and } T_3 = 24.15\,\mathrm{mV} \), enabling consistent and reliable symbol classification without errors for the given interval, owing to the transmitter's robust pulse-shaping capabilities and the receiver's fast, real-time response.

In contrast, Fig. \ref{fig:pwandcomm}(d) demonstrates the conditions in which ISI is introduced by reducing $T_\mathrm{g}$ to 3\,s and $T\textsubscript{pi}$ to 10\,s, causing significant overlap between consecutive pulses, leading to pulse dispersion and increased variance in amplitude and width. The detection process became more challenging, especially with smaller \(T_\mathrm{g}\), as slight flow-rate variations during ON/OFF transitions introduced additional timing misalignments and transposition errors. Although shorter $T_\mathrm{g}$ and $T\textsubscript{pi}$ enable higher transmission rates, this reduction comes at the cost of an increased risk of detection errors.  

\section{Conclusion}
\label{sec:Conclusions}

This study introduces a cost-effective, microfluidic-based, end-to-end MC testbed that integrates hydrodynamic gating for programmable pulse shaping and controlled signal transmission, and a screen-printed sensor for high-sensitivity detection. A pH-based MC system was implemented to investigate the testbed's capabilities. Hydrodynamic gating-based transmitter generated programmable pH-modulated MC signals by adjusting the flow rate ratio between the supply (pH 3) and gating (PBS, pH 7.4) solutions. Screen-printed pH sensor-based receiver, integrated into the microfluidic chip and connected to a readout board, enabled real-time MC signal detection. Pulse amplitude and width characterization confirmed that this low-cost, customizable testbed supports a wide range of MC scenarios with controlled pulse profiles. We conducted pH-based MC experiments under two conditions: with and without ISI. Sufficient pulse spacing enabled accurate 4-ary CSK classification, while shorter intervals introduced ISI, leading to signal distortion and detection challenges. The testbed can replicate various MC scenarios by adjusting inter-pulse duration, pulse widths and amplitudes. Future work could integrate multiplexed sensors, diversify channel geometries, and use biocompatible materials for \emph{in vivo} experiments.

\bibliographystyle{IEEEtran}
\bibliography{manuscript}

\begin{thebibliography}{10}
\providecommand{\url}[1]{#1}
\csname url@rmstyle\endcsname
\providecommand{\newblock}{\relax}
\providecommand{\bibinfo}[2]{#2}
\providecommand\BIBentrySTDinterwordspacing{\spaceskip=0pt\relax}
\providecommand\BIBentryALTinterwordstretchfactor{4}
\providecommand\BIBentryALTinterwordspacing{\spaceskip=\fontdimen2\font plus
\BIBentryALTinterwordstretchfactor\fontdimen3\font minus
  \fontdimen4\font\relax}
\providecommand\BIBforeignlanguage[2]{{%
\expandafter\ifx\csname l@#1\endcsname\relax
\typeout{** WARNING: IEEEtran.bst: No hyphenation pattern has been}%
\typeout{** loaded for the language `#1'. Using the pattern for}%
\typeout{** the default language instead.}%
\else
\language=\csname l@#1\endcsname
\fi
#2}}

\bibitem{akyildiz2015internet}
I.~F. Akyildiz, M.~Pierobon, S.~Balasubramaniam, and Y.~Koucheryavy, ``{The
  Internet of Bio-Nano Things},'' \emph{IEEE Commun. Mag.}, vol.~53, no.~3, pp.
  32--40, 2015.

\bibitem{kuscu2019transmitter}
M.~Kuscu, E.~Dinc, B.~A. Bilgin, H.~Ramezani, and O.~B. Akan, ``Transmitter and
  receiver architectures for molecular communications: A survey on physical
  design with modulation, coding, and detection techniques,'' \emph{Proc.
  IEEE}, vol. 107, no.~7, pp. 1302--1341, 2019.

\bibitem{lotter2301experimental}
S.~Lotter, L.~Brand, V.~Jamali, M.~Sch{\"a}fer, H.~M. Loos, H.~Unterweger,
  S.~Greiner, J.~Kirchner, C.~Alexiou, D.~Drummer, \emph{et~al.},
  ``Experimental research in synthetic molecular communications--part i,''
  \emph{IEEE Nanotechnol. Mag.}, vol.~17, no.~3, pp. 42--53, 2023.

\bibitem{lotter2023experimental}
------, ``Experimental research in synthetic molecular communications--part
  ii,'' \emph{IEEE Nanotechnol. Mag.}, vol.~17, no.~3, pp. 54--65, 2023.

\bibitem{angerbauer2023salinity}
S.~Angerbauer, M.~Hamidovic, F.~Enzenhofer, M.~Bartunik, J.~Kirchner,
  A.~Springer, and W.~Haselmayr, ``Salinity-based molecular communication in
  microfluidic channels,'' \emph{IEEE Trans. Mol. Biol. Multi-Scale Commun.},
  vol.~9, no.~2, pp. 191--206, 2023.

\bibitem{bhattacharjee2020testbed}
S.~Bhattacharjee, M.~Damrath, F.~Bronner, L.~Stratmann, J.~P. Drees,
  F.~Dressler, and P.~A. Hoeher, ``A testbed and simulation framework for
  air-based molecular communication using fluorescein,'' in \emph{Proc. ACM
  NANOCOM}, 2020, pp. 1--6.

\bibitem{walter2023real}
V.~Walter, D.~Bi, A.~Salehi-Reyhani, and Y.~Deng, ``Real-time signal processing
  via chemical reactions for a microfluidic molecular communication system,''
  \emph{Nature Communications}, vol.~14, no.~1, p. 7188, 2023.

\bibitem{kuscu2021fabrication}
M.~Kuscu, H.~Ramezani, E.~Dinc, S.~Akhavan, and O.~B. Akan, ``Fabrication and
  microfluidic analysis of graphene-based molecular communication receiver for
  {Internet of Nano Things (IoNT)},'' \emph{Scientific Reports}, vol.~11,
  no.~1, p. 19600, 2021.

\bibitem{naqvi2023testbed}
A.~H. Naqvi and B.~D. Unluturk, ``A testbed for molecular communication using a
  microfluidic device with fluorescent nanoparticles,'' in \emph{Proc. ACM
  NANOCOM}, 2023, pp. 144--145.

\bibitem{civas2023graphene}
M.~Civas, M.~Kuscu, O.~Cetinkaya, B.~E. Ortlek, and O.~B. Akan, ``Graphene and
  related materials for the internet of bio-nano things,'' \emph{APL
  Materials}, vol.~11, no.~8, 2023.

\bibitem{hamidovic2024microfluidic}
M.~Hamidovi{\'c}, S.~Angerbauer, D.~Bi, Y.~Deng, T.~Tugcu, and W.~Haselmayr,
  ``Microfluidic systems for molecular communications: A review from theory to
  practice,'' \emph{IEEE Trans. Mol. Biol. Multi-Scale Commun.}, 2024.

\bibitem{bolhassan2024microfluidic}
I.~M. Bolhassan, A.~Abdali, and M.~Kuscu, ``Microfluidic molecular
  communication transmitter based on hydrodynamic gating,'' \emph{IEEE Trans.
  Mol. Biol. Multi-Scale Commun.}, vol.~10, no.~1, pp. 185--196, 2024.

\bibitem{akyol2024experimental}
E.~Akyol, A.~B. Ozturk, I.~M. Bolhassan, and M.~Kuscu, ``Experimental
  characterization of hydrodynamic gating-based molecular communication
  transmitter,'' in \emph{Proc. ACM NANOCOM}, 2024, pp. 60--65.

\bibitem{farsad2016molecular}
N.~Farsad and A.~Goldsmith, ``A molecular communication system using acids,
  bases and hydrogen ions,'' in \emph{Proc. IEEE SPAWC}.\hskip 1em plus 0.5em
  minus 0.4em\relax IEEE, 2016.

\bibitem{franco2020screen}
F.~F. Franco, L.~Manjakkal, and R.~Dahiya, ``Screen-printed flexible carbon
  versus silver electrodes for electrochemical sensors,'' in \emph{Proc. IEEE
  FLEPS}.\hskip 1em plus 0.5em minus 0.4em\relax IEEE, 2020, pp. 1--4.

\bibitem{gvozdenovic2011electrochemical}
M.~M. Gvozdenovi{\'c}, B.~Z. Jugovi{\'c}, J.~S. Stevanovi{\'c}, T.~L.
  Tri{\v{s}}ovi{\'c}, and B.~N. Grgur, ``Electrochemical polymerization of
  aniline,'' \emph{Electropolymerization}, vol. 2011, pp. 77--96, 2011.

\end{thebibliography}

\vfill

\end{document}